# Growth of Nanosize and Colloid Particles by Controlled Addition of Singlets

Vladimir Privman
Center for Advanced Materials Processing,
Clarkson University, Potsdam, NY 13699-5820, U.S.A.

**ABSTRACT**

We outline a theoretical framework for estimating the evolution of the particle size distribution in colloid and nanoparticle synthesis, when the primary growth mode is by externally controlled addition of singlet building blocks. The master equations, analyzed in the leading "non-diffusive" expansion approximation, are reduced to a set of easily numerically programmable relations that yield information on the time evolution of the particle size distribution.

**INTRODUCTION**

Recently, we have developed a theoretical model [1,2] that can describe semiquantitatively, without adjustable parameters, growth of monodispersed colloids by precipitation from homogeneous solutions [3-8]. It has been established [1-8] that most monodispersed colloids, i.e., particles of narrow size distribution, with sizes ranging from submicron to few microns, consist of crystalline subunits. The latter are formed as primary particles in a supersaturated solution by the burst-nucleation mechanism [1,9,10]. They then further grow and aggregate, primarily by the mechanism of singlet capture by growing aggregates. At the same time, the resulting aggregates rapidly restructure to form compact, frequently spherical, polycrystalline secondary particles, of density close to that of the bulk material and of narrow size distribution.

This two-stage growth mechanism is illustrated in Figure 1. Many important theoretical issues have remained unsolved or only partially addressed in the literature. These include, in particular, shape selection [11-13] and particle morphology properties. Another important open topic involves extensions to nanosize particles. Such particles of sizes less than colloid, from few tens to single nanometers, involve several new interesting challenges and paradigms related to nanotechnology.

Firstly, what do we mean by "monodispersed" on the nanoscale? It is quite likely that for most truly large-molecule-dimension nanotechnology applications, uniform size (and shape) really means "atomically identical." This is particularly true for future electronic devices that involve quantum effects or quantum control. For many other applications, requirements for uniformity will be also quite strict. Therefore, methods of controlling size and shape distribution, which found numerous applications for colloids, will be even more important for nanotechnology.

In this work, we outline an evaluation method of particle growth controlled by addition of singlet building blocks. We hope that these ideas will initiate the application of some of the techniques developed in colloid science to nanoparticle synthesis. Our results will be also of interest in interpreting growth, specifically after seeding, in colloidal synthesis.

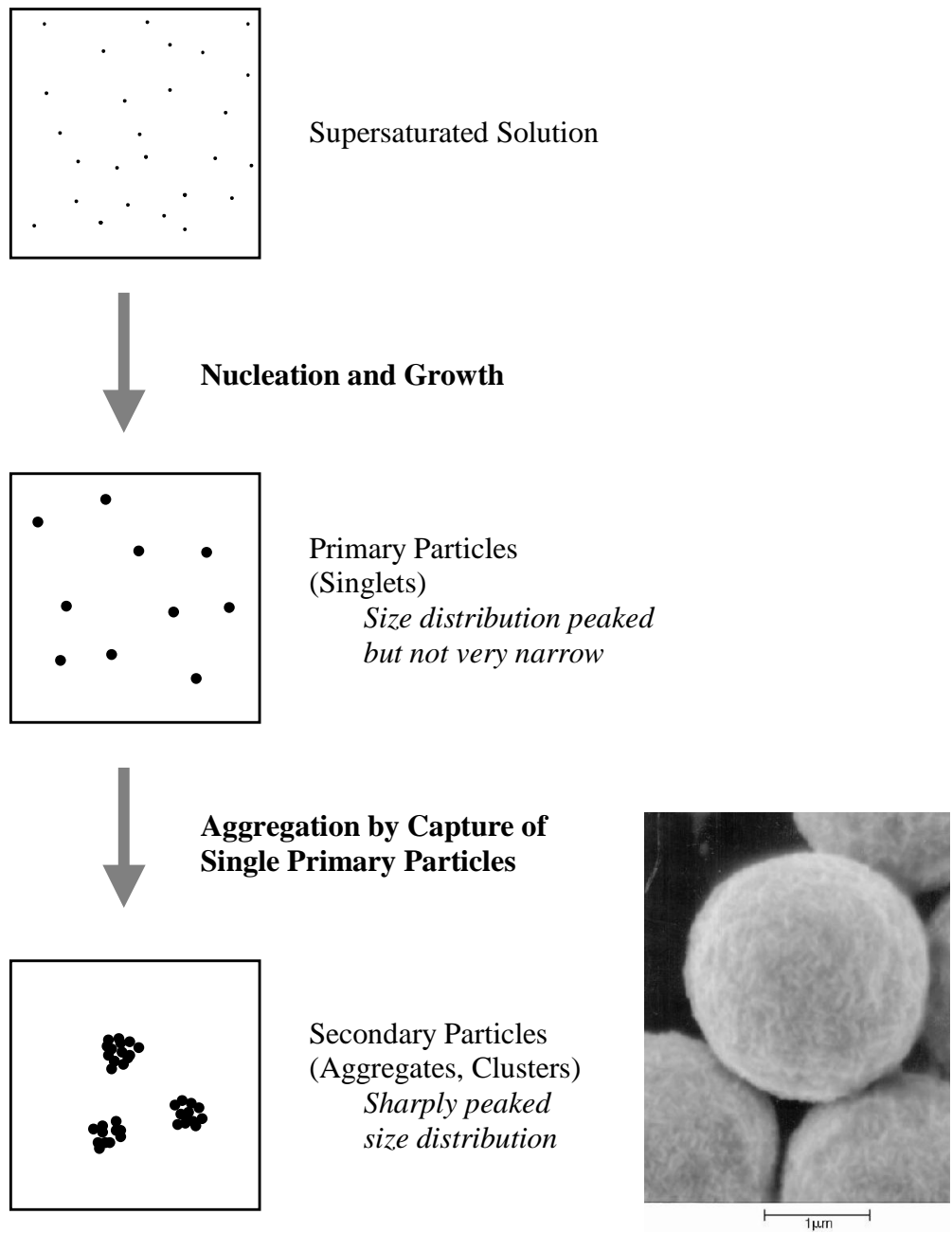

**Figure 1.** In colloid synthesis, initially, solutes are formed to yield a supersaturated solution, leading to nucleation. The formed nuclei may further grow by diffusive mechanism. The resulting primary particles (singlets) aggregate to form secondary particles. This latter process is sometimes facilitated by changes in the chemical conditions in the system: the ionic strength may increase, or the pH may change, causing the surface potential to approach the isoelectric point. Formation of the final (secondary) particles, of narrow size distribution, is a diffusion-controlled aggregation process, proceeding via the addition-polymerization type growth by irreversible capture of primary particles by the aggregates. At the same time, the aggregates also restructure into compact final particles, exemplified here by the SEM image of a gold colloid particle.

## GROWTH BY ADDITION OF SINGLETS

The singlet building blocks in nanoparticle synthesis in solution are atomic-size species (atoms, ions, molecules), whereas for colloid synthesis of the type described in Figure 1, they are the (nanosize) primary particles. Furthermore, in the process of Figure 1, the supply of singlets is "naturally" controlled by the features of their nucleation process. However, in principle the singlets can be added externally. In nearly any such process, the initial components are supplied over some interval of time. Their mixing in, must be fast enough to ensure uniform volume distribution. This represents an important practical chemical engineering problem.

In is therefore quite natural to consider the time dependence of the singlet addition, and its impact on the size distribution of the products. Specifically, for nanosize particle preparation, there has been recent interest in stepwise processes [14,15]: after achieving the initial particle distribution, batches of singlets are added to induce further growth.

Let $N_s(t)$ denote the volume density of particles, consisting of $s$ singlets, at time $t$. We are interested in the situation illustrated in Figure 2, when the particle size distribution evolves in time with a sharp peak eventually present at some rather large $s$ values. For convenience, let us denote the singlet concentration by

$$C(t) = N_1(t) \ . \tag{1}$$

The singlets can be supplied at the rate $\rho(t)$ per unit volume. They are consumed by the processes involving the production of the small clusters, in the "shoulder" in Figure 2. They are also consumed by the growing large clusters in the peak. There are two issues to consider in such growth: how is the peak created in the first place, and how to grow it without much broadening.

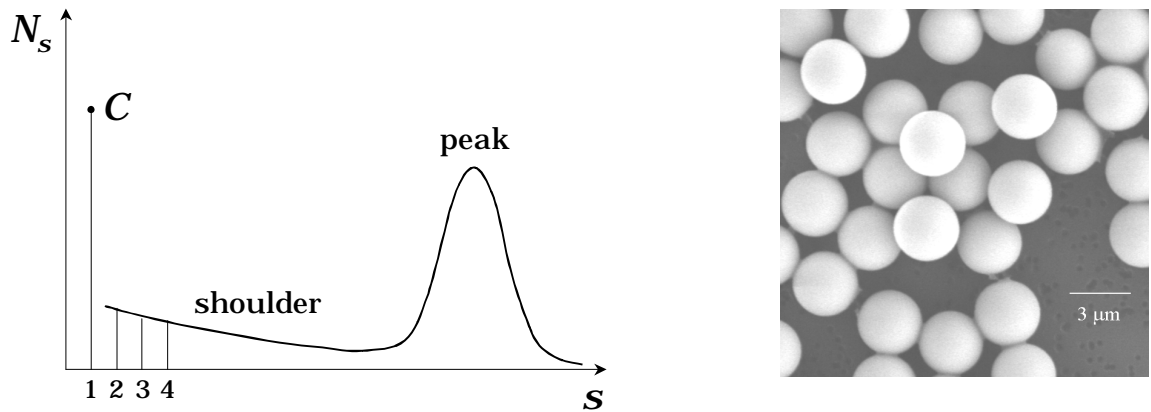

**Figure 2.** The expected form of the particle size distribution in uniform colloid or nanoparticle synthesis in solution. The peak at the large cluster sizes is growing at the expense of the supplied singlets, the concentration of which is controlled externally. The distribution for $s > 1$ can be usually assumed to be a smooth function of $s$, though the vertical bars at $s = 2, 3, 4$ emphasize that the $s$ values are actually discrete. The SEM image of cadmium sulfide colloid particles illustrates the attainable sharpness of the size distribution.

We will consider growth dominated by the irreversible capture of singlets by the larger growing aggregates. Thus, we use the rate equations, with $\Gamma_s$ denoting the rate constants for singlet capture by the $s \geq 1$ aggregates,

$$\frac{dN_s}{dt} = (\Gamma_{s-1} N_{s-1} - \Gamma_s N_s) C, \qquad s > 2, \qquad (2)$$

$$\frac{dN_2}{dt} = (\frac{1}{2}\Gamma_1 C - \Gamma_2 N_2) C, \qquad (3)$$

where $C = N_1$, as defined in (1), and

$$\frac{dC}{dt} = \rho - \sum_{s=2}^{\infty} s \frac{dN_s}{dt} = \rho - \Gamma_1 C^2 - C \sum_{s=2}^{\infty} \Gamma_s N_s. \qquad (4)$$

Let us point out that the assumption that the *only* process involving the $s > 1$ aggregates is that of capturing singlets at the rate proportional to the concentration of the latter, $\Gamma_s C$, is drastic but commonly used in the colloid literature, e.g., [1,2,12,16-18]. In fact, more complex processes, such as cluster-cluster aggregation [19,20], detachment [21] and exchange of singlets (ripening), etc., also contribute to particle growth. However, in uniform colloid synthesis they are typically much slower than the singlet-consumption growth. In addition, they broaden the particle size distribution.

Another important approximation involved in writing the relations (2)-(4) is that of ignoring particle shape distribution and their morphology. We avoid this issue, which is not well understood, by assuming that the growing aggregates rapidly restructure into compact bulk-like particles, of an approximately fixed shape, typically, but not always, spherical. This has been experimentally observed in uniform colloid particle synthesis [1,3-8]. Without such restructuring, the aggregates would be fractal [20-22].

For nanosize particle synthesis, the only assumption in the above summary that can be questioned is that of ignoring singlet detachment for the particles in the shoulder in Figure 2. Indeed, unlike colloid growth, which is fast and irreversible for all $s$ in such synthesis processes, the nanosize particle growth will be typically held back by a nucleation barrier [1,14]. During the late stage growth, that follows the initial nucleation burst [9,10], the barrier can be quite high. The distribution in the shoulder will approach the equilibrium Boltzmann form, governed by the excess free energy of the aggregate formation. It is interesting to note that this fast equilibration means that the singlets "stored" in the small, "shoulder" aggregates will be released and available for consumption by larger aggregates in the peak, provided the latter indeed primarily feed on "free" singlets.

If the singlets are supplied constantly, then the distribution, both for colloids and nanoparticles, will develop a large shoulder at small aggregates, with no pronounced peak at $s \gg 1$. If the supply is limited, then only small aggregates will be formed. An interesting recent observation in studies of colloid synthesis [1,2] has been that there exist "protocols" of singlet supply, at the rate $\rho(t)$ which is a slowly decaying, sometimes rather complicated function of time, that yield peaked size distributions at large times. Furthermore, the primary process summarized in Figure 1, naturally "feeds" the secondary process just at a rate like this.

For nanoparticle synthesis, the main mechanism of the early formation of the peak is by burst nucleation, when nuclei of sizes larger than the critical size form by growing over the nucleation barrier. Of course, *seeding* is another way of initiating the peaked size distribution both for colloid and nanosize growth.

**GROWTH OF THE PEAKED SIZE DISTRIBUTION**

Solution of the rate (master) equations (2)-(4) requires numerical approaches and is not particularly illuminating as to the nature of the particle growth. Therefore, we will introduce several additional assumptions which will allow us to go a long way in simplifying the problems in closed analytical form. The main idea is that, once the peak is formed after some transient time or by seeding, the particles in the peak are the main consumers of the available singlets.

This assumes that the singlet concentration is controlled by adding them externally [14,15]. For nanoparticle growth, the addition is at such a rate that the nucleation barrier remains high. The shoulder will then adjust to assume an approximately equilibrium shape, but the production of new larger, supercritical aggregates will be negligible. For colloid growth, the shoulder will also evolve, and new larger particles will be generated. However, if the number of larger aggregates is already significant, they will dominate the consumption of singlets.

Thus, to prevent generation of new small aggregates, we put

$$\Gamma_1 = 0, \tag{5}$$

which is an approximation. Furthermore, we will assume that $s$ is a continuous variable, since we are interested in $s \gg 1$, and that $s$ varies in the range

$$0 \leq s < \infty. \tag{6}$$

For calculations assuming singlet transport by diffusion, one can take the large $s$ Smoluchowski expression for the rates [21,23],

$$\Gamma_{s \gg 1} = A s^{1/3}, \tag{7}$$

where $A$ is a known constant. Note that $\Gamma_{s \gg 1}$ is proportional to the aggregate linear dimension (which includes the factor $s^{1/3}$) and singlet diffusion constant. Our results apply for general $\Gamma_s$.

Our last approximation is introduced in writing the continuous $s$ form of the master equations (2). We retain only the leading $s$ derivative, ignoring the "diffusive" second-derivative term (and higher-order terms). The consequences of this approximation, already used in the literature, e.g., [12], will be discussed later. Thus, we replace (2) by

$$\frac{\partial N(s,t)}{\partial t} = -C(t) \frac{\partial}{\partial s} [\Gamma(s) N(s,t)], \tag{8}$$

with (4) replaced by

$$\frac{dC(t)}{dt} = \rho(t) - C(t) \int_0^\infty ds \left[ \Gamma(s) N(s,t) \right]. \tag{9}$$

Let us now define the variable [1,16,18]

$$\tau(t) = \int_0^t dt' C(t') \geq 0, \tag{10}$$

and then introduce the function $u(s,\tau)$ via the relation

$$\tau = \int_u^s \frac{ds'}{\Gamma(s')}. \tag{11}$$

We point out that usually $\Gamma(s) > 0$, and the lower limit of integration can be taken to zero. The asymptotic rate expression (7) does vanish at $s = 0$ because of our cavalier treatment of the small $s$ behavior. However, the integral happens to converge, so no additional care is needed. We can safely define the quantity $s_{\min}(\tau)$ via

$$\tau = \int_0^{s_{\min}} \frac{ds'}{\Gamma(s')}. \tag{12}$$

As $u$ is increased from zero to infinity, the corresponding $s(u,\tau)$, for fixed $\tau$, increases from $s_{\min}(\tau)$ to infinity.

Next, we notice that the relation between the differentials implied by (11), namely,

$$d\tau = \frac{ds}{\Gamma(s)} - \frac{du}{\Gamma(u)}, \tag{13}$$

allows us to calculate various partial derivatives in terms of $\Gamma(s)$ and $\Gamma(u) = \Gamma\big(u(s,\tau(t))\big)$. This, in turn, allows one to verify, by a cumbersome calculation not reproduced here, that (8) is solved by

$$N(s,t) = \frac{\Gamma\big(u(s,\tau(t))\big)}{\Gamma(s)} N\big(u(s,\tau(t)),0\big), \qquad s \geq s_{\min}\big(\tau(t)\big), \tag{14}$$

$$N(s,t) = 0, \qquad 0 \leq s \leq s_{\min}\big(\tau(t)\big), \tag{15}$$

where the discontinuity at $s_{\min}(\tau(t))$ is possible if the initial distribution at time zero, $N(s,0)$, is nonzero at $s=0$. Actually, within the present approximation of ignoring the effects of the details of the size distribution for small $s$, we could as well require that $N(0,0)=0$.

Let us summarize the above observations by emphasizing that we consider a particle size distribution which at time $t=0$ already has a well-developed significant peak at large cluster sizes. Relations (14)-(15) will provide an approximate description of the evolution of this peak with time, owing to supply of singlets at the rate $\rho(t)$. The form of the distribution at small particle sizes plays no role in the derivation.

In fact, neglecting the second-derivative in $s$, "diffusive" term in writing (8), leads to certain artificial conclusions. Specifically, sharp features and discontinuities of the initial distribution (as well as its derivatives, etc.) will not be smoothed out. The fact that the initial distribution is only meaningful for $s \geq 0$ translates into the sharp cutoff at $s_{\min}$ for times $t > 0$. Had we included the diffusive term, the distribution would extend smoothly to $s = 0$ for all times. However, no analytical solution (14)-(15) would be available.

While this lack of smoothness is probably not important for a semiquantitative evaluation of the size distribution, one aspect should be emphasized as critical: if the initial distribution is already very sharp, then the neglect of the diffusive term in our expressions may result in underestimating the *width* of the evolving peak.

To complete the description of the particle size distribution within the non-diffusive approximation, we have to discuss the estimation of the function $\tau(t)$. This is taken up in the next section, where we also further mention the matter of the non-diffusive approximation.

**EVALUATION OF THE PEAKED PARTICLE SIZE DISTRIBUTION**

Relations (9)-(10) can be rewritten, using (14), as a system of coupled differential equations for two unknown functions $\tau(t)$ and $C(t)$, with $\tau(0) = 0$, and $C(0)$ externally controlled,

$$\frac{d\tau}{dt} = C(t), \tag{16}$$

$$\frac{dC}{dt} = \rho(t) - C(t) F(\tau), \tag{17}$$

where

$$F(\tau) = \int_{s_{\min}(\tau)}^{\infty} ds \left[ \Gamma(u(s,\tau)) N(u(s,\tau),0) \right]. \tag{18}$$

These equations are easily programmed for numerical evaluation, especially if the function $F(\tau)$ is calculable analytically, so that numerical integration can be avoided. The latter is likely for the power-law rate in (7), provided the initial distribution $N(s,0)$ is not too complicated.

Within the approximation developed here, the number of particles larger than singlet, $M$, obviously remains constant,

$$M = \int_{s_{\min}(t)}^{\infty} ds N(s,t) = \int_{0}^{\infty} ds N(s,0). \tag{19}$$

The change in the average size of the particles larger than singlet,

$$\langle s \rangle_t = \frac{1}{M} \int_{s_{\min}(t)}^{\infty} ds [sN(s,t)], \tag{20}$$

can be evaluated directly from $C(t)$,

$$\langle s \rangle_t = \langle s \rangle_0 + \frac{1}{M}[C(0) - C(t) + \int_{0}^{t} dt' \rho(t')]. \tag{21}$$

Furthermore, consideration of the increment relations following from (13), suggests that the width of the peak, $W_t$, grows according to

$$W_t \approx \frac{\Gamma(\langle s \rangle_t)}{\Gamma(u(\langle s \rangle_t, \tau(t)))} W_0 > W_0. \tag{22}$$

The inequality follows from the definition (11), assuming that for large $s$, $\Gamma(s)$ is an increasing (positive) function. This excludes an important case of constant $\Gamma$, appropriate for certain models of polymerization. In this case, however, the discrete equations (2)-(4) can be analyzed directly in great detail [16,18], so that the present formulation is not needed.

 In connection with (22), the reader must be cautioned that additional broadening will result from the second-derivative "diffusive" term neglected in our continuous $s$ equations. The model with the diffusive term included, requires serious numerical efforts, as does the original, discrete $s$ model [1,2].

 In summary, with the reservations regarding the width estimates, numerical calculation of the functions $\tau(t)$ and $C(t)$, via (16)-(18), goes a long way in estimating various properties of the growing, peaked particle size distribution. Results and applications, specifically for the case of the Smoluchowski rate (7), and comparison with large-scale simulation results of the discrete $s$ equations of the type (2)-(4), will be reported in forthcoming publications.

 Even at the level of the approximation (22), it is obvious that the size distribution never actually narrows in absolute terms. All the experimentally realized monodispersed particle synthesis procedures in solution, investigated thus far primarily in the colloid domain, actually yield small *relative* peak width, $W_t / \langle s \rangle_t$, by utilizing fast increase in $\langle s \rangle_t$ via consumption of singlets, on the time scales short for the "diffusive" broadening to set in.

**ACKNOWLEDGEMENTS**

The author gratefully acknowledges helpful discussions and collaboration with M. Lawrence Glasser, Dan Goia, Sergiy Libert, Egon Matijevic, Dima Mozyrsky, Jongsoon Park and Yitzhak Shnidman. This research has been supported by the National Science Foundation (grant DMR-0102644) and by the Donors of the Petroleum Research Fund, administered by the American Chemical Society (grant 37013-AC5,9).